\pdfoutput=1
\documentclass[a4paper]{jpconf}
\usepackage{graphicx}

\begin{document}
\title{{\sc MadAnalysis}~5: status and new developments}

\author{\underline{Eric Conte}$^1$ and Benjamin Fuks$^{2,3}$}
\address{
  $^{(1)}$ Groupe de Recherche de Physique des Hautes \'Energies (GRPHE),
       Universit\'e de Haute-Alsace, IUT Colmar, 34 rue du Grillenbreit BP 50568,
       68008 Colmar Cedex, France\\
  $^{(2)}$ Theory Division, Physics Department, CERN, CH-1211 Geneva 23, Switzerland\\
  $^{(3)}$ Institut Pluridisciplinaire Hubert Curien/D\'epartement Recherches Subatomiques, 
    Universit\'e de Strasbourg/CNRS-IN2P3, 23 Rue du Loess, F-67037 Strasbourg, France}
\ead{eric.conte@iphc.cnrs.fr, benjamin.fuks@iphc.cnrs.fr}

\begin{abstract}
\textsc{MadAnalysis~5} is a new {\sc Python}/{\sc C++} package facilitating
  phenomenological analyses that can be performed in the framework of Monte Carlo simulations of collisions to be
  produced in high-energy physics experiments. It allows one, by means of a user-friendly
  interpreter, to perform professional physics analyses in a very simple way. Starting from
  event samples as generated by any Monte Carlo event generator, large classes of selections
  can be implemented through intuitive commands, many standard kinematical distributions can
  be automatically represented by histograms and all results are eventually
  gathered into detailed {\sc Html} and \textsc{latex}
  reports. In this work, we briefly report on the latest developments of the code, focusing
  on the interface to the \textsc{FastJet} program dedicated to jet reconstruction.
\end{abstract}

\section{Introduction}


The TeV energy regime is currently being explored at the Large Hadron Collider, at CERN, and searches for
new phenomena are a central topic of the associated experimental program. However,
as a consequence of the absence of hints for new physics,
the states predicted by any theory beyond the Standard Model (BSM)
are bound to lie at higher and higher scales. While experimental
analyses mostly focus on popular BSM theories and on simplified extensions
of the Standard Model (SM), (re)interpretations of the results in any specific theoretical framework
and phenomenological investigations of new types of signatures play key roles
in the improvement of our understanding of the TeV scale.
To this aim, matrix-element-based Monte Carlo event generators
are generally employed, the results being matched to parton showering, hadronization and detector simulation algorithms.
Phenomenological analyses can then be performed at several levels, from a purely partonic description
of the simulated event final states to a more complex description including hadronization
and possibly a fast simulation of the detector response.
The \textsc{MadAnalysis~5} package~\cite{MA5} offers a unique framework to perform such analyses, in an easy and efficient way.
While about to be fully embedded within the Monte Carlo event generators \textsc{MadGraph}~5~\cite{MG5} and aMC@NLO~\cite{aMCNLO},
\textsc{MadAnalysis~5} can also be used as a standalone program and outside the {\sc MadGraph} environment.

In this paper, we describe some of the novel features of the version 1.1.8 of the code
and provide an update of the original user guide. After a brief overview of
{\sc MadAnalysis}~5 in Section~\ref{sec:overview}, we illustrate how to use it to
design a phenomenological analysis in Section~\ref{sec:stepbystep}.
Section~\ref{sec:fastjet}
is dedicated to a recent module added the code, related to jet reconstruction and allowing one to
verify the merging procedures usually employed to combine event samples describing the same physics process but
with different final state jet multiplicities.
Our conclusions are presented in Section~\ref{sec:conclusion}.

\section{Overview}\label{sec:overview}

\subsection{\textsc{MadAnalysis~5} in a nutshell}

\textsc{MadAnalysis~5} is a framework designed to facilitate the implementation of
phenomenological studies based on Monte Carlo simulations of collisions describing both new physics
signal and Standard Model background processes.
It allows one, on the one hand, to investigate the characteristics of a particular process,
and, on the other hand, to carry out an event selection aiming to extract a given signal
from the background. It supports event files encoded under the {\sc StdHep}~\cite{STDHEP},
{\sc HepMc}~\cite{HEPMC}, {\sc Lhe}~\cite{LHE1,LHE2}, {\sc Lhco}~\cite{LHCO} and
\textsc{Delphes Root}~\cite{DELPHES} formats.

Properties of the event final states can be probed by means of a small set of intuitive {\sc Python}
commands (see below) acting
on particles modeled by user-defined labels and several observables. They can then be
represented by histograms whose layout can be tuned according to the preferences of the user. In the same
way, event selection criteria can be implemented via dedicated commands and equally applied
to a full event (\textit{i.e.}, to select or reject it under a given condition) or
to a particle (\textit{i.e.}, to include it or not in the analysis under
a specific condition). {\sc MadAnalysis}~5 can also further compute signal over background
ratios whose definition can be adjusted by the user.

Flexible, \textsc{MadAnalysis~5} also offers sophisticated functionalities allowing one
to, \textit{e.g.}, combine particle four-momenta (sums, differences, \textsl{etc.}) before the
computation of any observable,
use the event history (mother-to-daughter relations linking a particle to its decay products)
or identify a given particle according to its hardness.
All the results are collected into reports available under the \textsc{Html}, \textsc{Postscript}
(associated with \textsc{latex}) or \textsc{Pdf} (associated with \textsc{pdflatex}) formats.
Interfaced to the \textsc{FastJet} package~\cite{FASTJET}, \textsc{MadAnalysis~5}
also includes several jet-clustering algorithms that can be used for object reconstruction,
including or not a simplified detector simulation and several (mis)identification efficiencies.

\subsection{Python console}

Once downloaded from the Internet (see Ref.~\cite{MA5WEBSITE}) and unpacked, the \textsc{MadAnalysis~5} package
does not require any compilation and can be started by issuing
\begin{verbatim}
  bin/ma5
\end{verbatim}
from the directory where it has been stored. By default, \textsc{MadAnalysis~5} is configured for
the analysis of events at the parton level,
although events at the hadron or at the reconstructed level can
be analyzed by launching the program with the \texttt{-H} and \texttt{-R} arguments, respectively.

The user interacts with \textsc{MadAnalysis~5} through a console
implemented in the \textsc{Python} language that invites him/her to define his/her analysis
through a set of commands that can also possibly be collected into a script. The \textsc{MadAnalysis}~5
metalanguage is designed to be simple and concise, as illustrated in the examples presented
in the next sections, and the user can be helped in two ways. First, in-line help can be
employed by issuing
\begin{verbatim}
  ma5> help
\end{verbatim}
in the console, which displays the list of available commands. Second, auto-completion has been implemented and can be used via the
\textit{tab} key.

\subsection{Structure of the program}

\begin{figure}[!h]
\centering
\includegraphics[width=.6\textwidth]{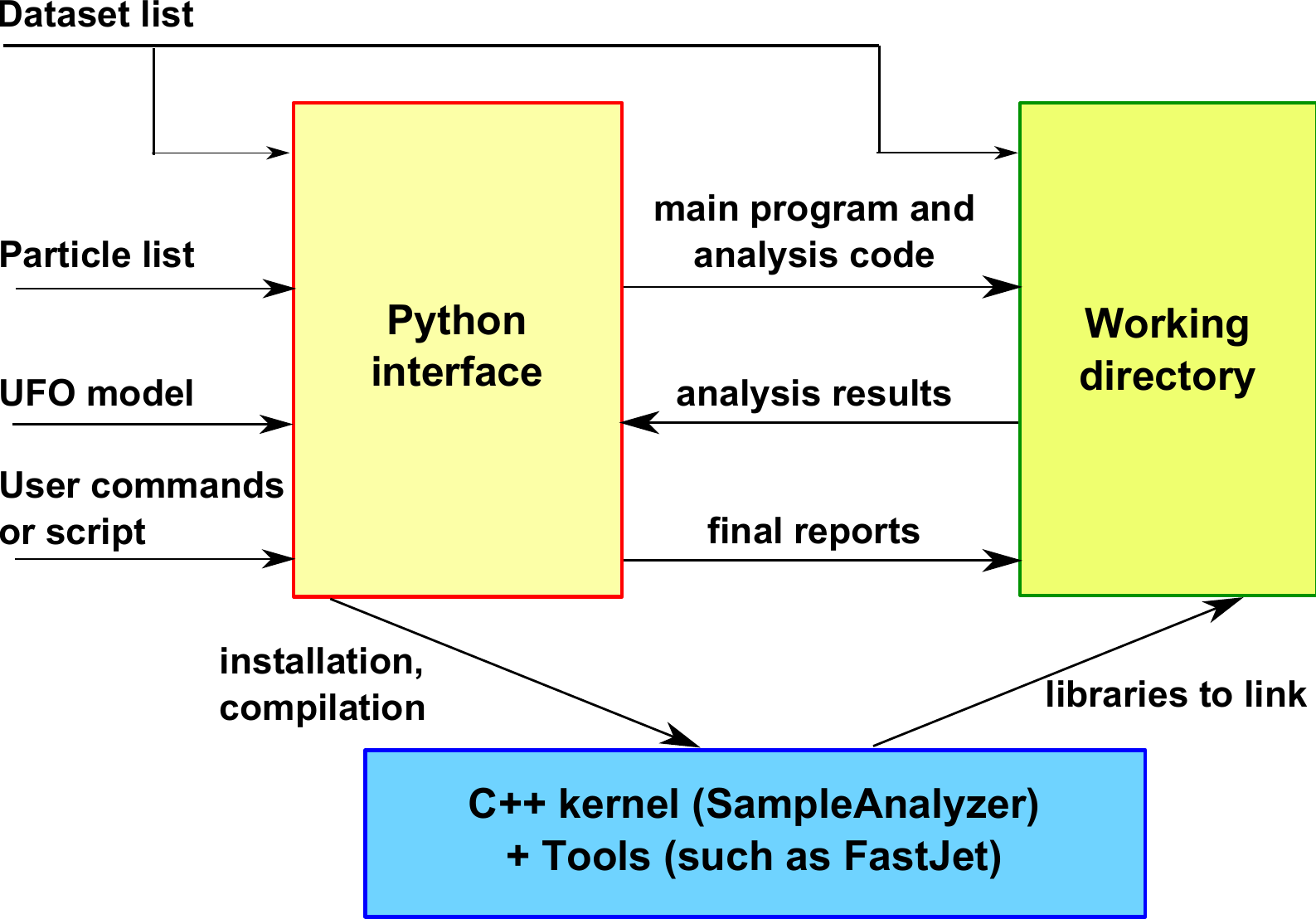}
\caption{Logical structure of \textsc{MadAnalysis~5} \label{fig:structure}}
\end{figure}

Particle labels can be either directly defined in the \textsc{Python} console or imported
from a Universal FeynRules Output (UFO) model containing names and labels for all the particles
of a given model~\cite{UFO}.
Additionally, the list of event samples to analyze has to be specified.
{\sc MadAnalysis}~5 then takes care of automatically exporting the analysis under the form
of a {\sc C++} code to be linked to a library called \textsc{SampleAnalyzer}, shipped with the
package, and to any other optional tools such as \textsc{FastJet} or \textsc{Delphes}
according to the needs of the user. The {\sc C++} program is further compiled and executed, and
the results are encapsulated into an {\sc Xml} structure stored in a so-called {\sc Saf} (standing
for \textsc{SampleAnalyzer} Format) file. The \textsc{Python} console is eventually automatically
taking care of rendering
the information human-readable, under the form of \textsc{PostScript}, \textsc{Pdf} and \textsc{Html} reports including
the defined histograms and the computed selection efficiencies.
The full code structure is summarized on Figure~\ref{fig:structure}.

On different footings, users can also directly implement their analysis in the {\sc C++} language,
bypassing the \textsc{Python} console and using all built-in functions available within
the {\sc SampleAnalyzer} library. This way of using \textsc{MadAnalysis~5} is called the \textsl{expert mode}
of running, in contrast to the \textit{normal mode} using the {\sc Python} console,
and allows one to implement any analysis which a user can dream of,
the only limitation being his/her imagination and programming skills.

\subsection{Requirements}
\label{sec:install}
When started, \textsc{MadAnalysis~5} firstly checks the presence on the system of required dependencies and
secondly verifies which of optional libraries are available. The list of the mandatory packages to run properly the program is the following~:
\begin{itemize}
  \item \textsc{Python} 2.6 or a more recent version (although not a 3.x version),
  \item the {\sc Gnu Gcc} compiler together with the associated Standard Template Libraries,
  \item the \textsc{Root} package version 5.27 or more recent~\cite{Brun:1997pa}.
\end{itemize}

Some specific functionalities of \textsc{MadAnalysis~5} may require the use of
other external packages and are thus deactivated if necessary.
Those consist of
\begin{itemize}
  \item the \textsc{latex} and \textsc{pdflatex} compilers (analysis reports in the \textsc{PostScript} and {\sc Pdf} formats,
    respectively),
  \item the \textsc{zlib} package (reading and writing compressed files),
  \item the \textsc{FastJet} package (to switch on the module dedicated to jet reconstruction, see Section~\ref{sec:fastjet}),
  \item the \textsc{Delphes} package (reading {\sc Root} files as outputted by {\sc Delphes}~3).
  \item a few representative event samples useful to learn how to use \textsc{MadAnalysis~5}.
\end{itemize}

With the exception of the \textsc{latex} and \textsc{pdflatex} compilers (that must
be installed independently by the user), each optional package that {\sc MadAnalysis}~5 may require
can be installed automatically by using the command \texttt{install} from the \textsc{Python} console.
For instance, typing in the commands
\begin{verbatim}
  ma5> install gzip
  ma5> install delphes
  ma5> install fastjet
  ma5> install samples
\end{verbatim}
leads to the installation of the {\sc zlib} libraries, the {\sc FastJet} and {\sc Delphes}~3 programs
and some Monte Carlo event samples, respectively.

\section{Designing an analysis with \textsc{MadAnalysis~5}}
\label{sec:stepbystep}
In this section, we briefly describe the way to perform a simple analysis with \textsc{MadAnalysis~5}. More advanced functionalities as well as the expert mode of the code
are not discussed and we refer to the manual for more details~\cite{MA5}.

\subsection{Basic concepts}

First, the event samples that must be read and analyzed
are specified through the command \texttt{import}. By default, all imported samples are collected within
a single dataset called \texttt{defaultset}, although the user can define as many datasets as necessary
by using the command \texttt{import} together with the keyword \texttt{as}.
In general, it is recommended to have one single dataset for each physics process.
The wildcard characters \texttt{*} and \texttt{?} can
also be used to import several files simultaneously. 

Second, in order to treat the different particle species all along the analysis,
\textsc{MadAnalysis~5} employs labels referring either to one given type of particle or to several types of
particles (the term of multiparticle is used in this case). When \textsc{MadAnalysis~5} starts, collections of standard labels for the SM and Minimal Supersymmetric Standard Model particles are loaded.
The user has then the possibility to define his/her own labels by the command \texttt{define}.

Third, histograms representing either properties of the full event (\textit{e.g.}, the transverse
missing energy or the transverse hadronic energy) or properties
of a given particle assumed to be part of the event final state content (\textit{e.g.},
its transverse momentum or its multiplicity) can be created by using
the command \texttt{plot}, followed by the name of the relevant observable
(see the manual for more details on the syntax~\cite{MA5}).

Finally, the list of observables that can be represented by histograms can
also be used to apply selection requirements either on the events themselves, or
to select the final state objects to be
considered in the analysis. To this aim, the user can employ two complementary commands, \texttt{reject} and \texttt{select}, preceding a criterion.

\subsection{Illustrative example}

We provide, for the sake of the example, a toy analysis based on the events samples provided by installing
the optional package \textsc{samples}. The analysis commands are given by
\begin{verbatim}
  (1) ma5> import  samples/zz.lhe.gz  as  mySignal
  (2) ma5> import  samples/ttbar_sl_?.lhe.gz  as  myBkg
  (3) ma5> set  myBkg.type  =  background
  (4) ma5> define  mu  =  mu+  mu-
  (5) ma5> plot  MET
  (6) ma5> plot PT(mu) [logY]
  (7) ma5> reject  (mu)  PT  <  20
  (8) ma5> select  80  <  M  (mu+  mu-)  <  100
  (9) ma5> plot  M(mu+  mu-)
\end{verbatim}

On the first line, a dataset called \texttt{mySignal} is created and linked to the event file \texttt{zz.lhe.gz}. On the second line, one collects the two samples \texttt{ttbar\_sl\_1.lhe.gz} and \texttt{ttbar\_sl\_2.lhe.gz} into a unique dataset named
\texttt{myBkg}, which is further defined as containing background events on the third line. In contrast,
the type of the \texttt{mySignal} dataset is left to its default value, \textit{i.e.}, the related
events being considered as signal events.

A new multiparticle label is defined on the fourth line by making use of the command \texttt{define} and
is associated with both muons and antimuons. This label is then used in the rest of the analysis
to define selections and histograms. Two histograms are hence defined on the fifth and sixth
lines of the analysis, the first one representing the missing transverse energy distribution extracted
from each dataset and the second one describing the transverse momentum
spectrum obtained after considering all muons and antimuons of
each event. By the commands of the seventh and eighth lines, we demand to consider, in the rest of the
analysis, all (anti)muons
with a transverse momentum smaller than 20~GeV and to keep events containing at least one
muon-antimuon pair with an invariant mass included in the range $[80, 100]$~GeV. A final histogram
is created on the last line of the analysis and allows one to study the effect of the previous
selections on the invariant mass of all muon-antimuon pairs that can be formed in each event.

\subsection{Submitting and executing the analysis}

\begin{figure}[!h]
\centering
\includegraphics[width=.75\textwidth]{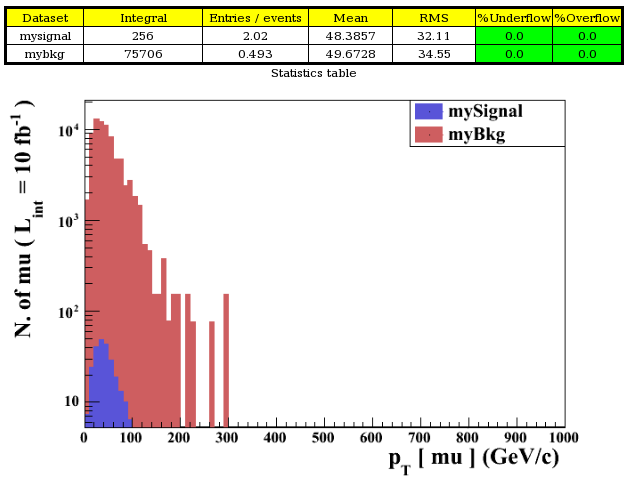}
\begin{tabular}{p{2cm}l}
  & \vspace*{-11cm}\\
  & \texttt{ma5> plot PT(mu) [logY]}
\end{tabular}
\vspace*{-0.7cm}
\caption{Histogram representing the transverse momentum spectra of all (anti)muons for the two datasets denoted by \texttt{mySignal} and \texttt{myBkg}. The command associated with the histogram generation
is superimposed to the figure.
  \label{fig:report}}
\end{figure}

The user can execute his/her analysis by typing in, in the {\sc Python} console,
the command
\begin{verbatim}
  submit
\end{verbatim}
Behind the scenes, a {\sc C++} code corresponding to the implemented analysis
is generated, compiled and linked to the
\textsc{SampleAnalyzer} library. The executable is then run over all datasets and the results are
displayed via the different reports. The {\sc Html} report can always be open directly from the {\sc Python} interface by means of the command
\begin{verbatim}
  open
\end{verbatim}
A representative histogram that can be found in the report associated with the analysis above is shown in Figure~\ref{fig:report}.

\section{Interface to \textsc{FastJet} package}
\label{sec:fastjet}
In order to enable the functionalities of {\sc MadAnalysis}~5 related to jet clustering, the
{\sc FastJet} package has to be installed on the system of the user, either externally to {\sc MadAnalysis}~5
or within the program (see Section~\ref{sec:install}).

\subsection{Reconstructing jets in hadron-level events}

Starting from an event file at the hadron level (in general encoded in the {\sc StdHep} or {\sc HepMc} file format),
\textsc{MadAnalysis~5} offers the user a straightforward way to gather the hundreds of hadrons usually
present in the event final states into jets by means of any of the algorithms included
in the {\sc FastJet} library. After this procedure, the events are described
in terms of jets, electrons, muons, taus, photons and missing transverse energy. While leptonically-decaying
taus are accounted for in terms of their decay products (electrons or muons and missing energy),
only hadronically-decaying taus are considered as a distinct physics objects (and defined as taus).

In order to enable jet clustering within {\sc MadAnalysis}~5 (assumed to be started in the reconstructed-level mode),
it is sufficient to specify the algorithm to be used as the value of
the \texttt{main.clustering.algorithm} variable. The allowed choices are \texttt{kt}
(longitudinally-invariant $k_T$ algorithm~\cite{jetalgo:kt,jetalgo:kt2}), \texttt{cambridge} (the Cambridge/Aachen algorithm~\cite{jetalgo:cambridge,jetalgo:cambridge2}),
\texttt{antikt} (the anti-$k_T$ algorithm~\cite{jetalgo:antikt}), \texttt{genkt} (the generalized $k_T$ algorithm ~\cite{FASTJET}),
\texttt{siscone} (the {\sc SisCone} algorithm~\cite{jetalgo:siscone}), \texttt{cdfmidpoint} (the CDF {\sc MidPoint} algorithm~\cite{jetalgo:cdfmidpoint}),
\texttt{cdfalgo} (the CDF {\sc JetClu} algorithm~\cite{jetalgo:cdfjetclu}) and \texttt{gridjet} (the {\sc GridJet} algorithm~\cite{FASTJET}).
For instance, employing the anti-$k_T$ algorithm can be enforced by typing\begin{verbatim}
  ma5> set main.clustering.algo = antikt
\end{verbatim}
As soon as a given jet algorithm is selected, new options depending on the algorithm
parameters become
available. In the (illustrative) example of the anti-$k_T$ algorithm, the variables
\texttt{main.clustering.radius} and \texttt{main.clustering.ptmin}, respectively representing
the jet-radius parameter and the jet transverse-momentum threshold, can be set by the user.

We now turn to the description of all other options coming with the jet
clustering module of {\sc MadAnalysis}~5 and that are related to the
identification of the physics objects. After the application of the algorithm,
each particle is uniquely identified and cannot contribute to several collections of physics objects.
A remark on leptons is however in order here. By default, the jet algorithm distinguishes leptons issued from the hard process
from those originating from hadron decays. While the first category of leptons are reconstructed as leptons
and included in the relevant collection,
leptons of the second category are feeding the jet algorithm and therefore included as parts of the reconstructed jets.
This default behavior can be modified by setting the variable \texttt{main.clustering.exclusive} to \texttt{false}.
This forces the jet algorithm to include all visible particles for jet reconstruction,
so that a given lepton can consequently be part of
the relevant lepton collection and also contribute to a specific jet.

{\sc MadAnalysis} also includes a simple algorithm dedicated to the $b$-tagging of the reconstructed
jets, so that a jet is identified as originating from a $b$-quark
when it can be matched to a $b$ (anti)quark, \textit{i.e.},
when it lies within a cone of radius $R$ around one of the parton-level $b$ quarks. The size of the
cone ($R$) is specified as the value of the \texttt{main.clustering.bjet\_id.matching\_dr} variable,
while setting the parameter \texttt{main.clustering.bjet\_id.exclusive} to \texttt{true}
indicates the code that one cannot match several jets to one given $b$ quark. By default, the $b$-tagging
algorithm is perfect, although (mis)identification efficiencies can also be specified. For example,
typing the three commands
\begin{verbatim}
  ma5> set main.clustering.bjet_id.efficiency = 0.60 
  ma5> set main.clustering.bjet_id.misid_cjet = 0.10
  ma5> set main.clustering.bjet_id.misid_ljet = 0.05 
\end{verbatim}
first sets the $b$-tagging efficiency to 60\%. Next, the two other commands
teach the $b$-tagging algorithm that $c$-quark-initiated jets and other lighter jets are tagged
as $b$-jets in respectively 10\% and 5\% of the cases. If more advanced $b$-tagging methods
are required by the user,
such as implementing a tagging algorithm depending
on the jet transverse momentum, he/she is recommended to use instead
a program simulating the entire detector response such as {\sc Delphes}~\cite{DELPHES}.
Furthermore, similar options are available for hadronically-decaying taus.

\subsection{Storing jet-clustered events}

Events obtained after jet clustering can be stored in a file whose name is specified by
the user as the value of the variable \texttt{main.outputfile}. For instance,
\begin{verbatim}
  ma5> set main.outputfile = "myOutput.lhe"
\end{verbatim}
allows one to save the reconstructed events in a file named \texttt{myOutput.lhe}.
The file format is directly specified by the file extension, the suitable choices being either \texttt{lhe} ({\sc Lhe}
format) or \texttt{lhco} ({\sc Lhco} format). Compressed files can also be used
if the optional \textsc{zlib} libraries are installed, the related file extensions being in
this case \texttt{lhe.gz} and \texttt{lhco.gz}. The files created by {\sc MadAnalysis}~5 are however not
strictly following the {\sc Lhe} and {\sc Lhco} conventions originally defined
in Refs.~\cite{LHE1,LHE2,LHCO}.
\begin{itemize}
  \item Some variables associated with
    an {\sc Lhco} event, such as the calorimeter energy deposits, are ignored by {\sc MadAnalysis}~5.
  \item In the case of {\sc Lhe} events, initial state particles are recorded with a status
    code set to $-1$, final state particles with a status code equal to $1$ and intermediate
    particles involved in the hard process with a status code set to 3. Other intermediate particles, such
    as those generated by parton showering, are omitted from the output file.
    Each reconstructed physics object,  \textit{i.e.},
    each of the light jets, $b$-tagged jets,
    electrons, muons, hadronically-decaying taus, photons as well as the missing transverse energy, is identified with
    a specific code inspired by the Particle Data Group (PDG) particle numbering scheme, chosen to be
    21, $\pm5$, $\pm$11, $\pm$13, $\pm$15, 22 and 12, respectively. Finally, the link between an
    intermediate particle of the hard process and
    the induced final state objects is saved
    by means of the associated mother-to-daughter relations as they can be encoded in the {\sc Lhe} format.
\end{itemize}
We refer to the header of the outputted files for more details.

\subsection{Validating multiparton matrix-element merging}

In order to achieve an accurate description of the jet properties
of a given event sample, two complementary approaches are usually combined. Matrix elements are
employed for the description of the hard jets and parton showering for the one of the soft and collinear jets.
Consequently, a double counting of the QCD emission is possible as radiation can now emerge from two sources.
Several merging procedures have been subsequently
developed in the last decades. {\sc MadAnalysis}~5 offers
a way, based on its interface with {\sc FastJet}, to validate the resulting \textit{merged} samples.
This relies on a check of the
smoothness of the differential jet rate (DJR) distributions, the DJR observables
strongly depending on the merging parameters and procedure as they describe the distribution of the
scale at which the jet multiplicity of an event final state
switches from $N$ to $N+1$.
The DJR histograms can be included in the \textsc{MadAnalysis~5} reports by typing
\begin{verbatim}
  ma5> set main.merging.check = true
\end{verbatim}

\section{Summary}
\label{sec:conclusion}
\textsc{MadAnalysis~5} is a unique, flexible and user-friendly
framework allowing for professional analyses of
simulations of collisions to be produced in high-energy physics experiments.
It can be used equivalently for the investigation of parton-level,
hadron-level or reconstructed events and is therefore able
to read many input formats. The program comes with two modes of running, the so-called normal and expert modes.
In the first case, the user is invited to define his/her analysis by typing command lines in
a \textsc{Python} console. In the second case, he/she
can directly implement his/her analysis in {\sc C++}, using a large class
of predefined functions included in the core of the package.

In this paper, we have emphasized new functionalities related to an interface with the \textsc{FastJet}
program and which include the possibility of using different
jet algorithms to cluster hadrons into jets, when analyzing hadron-level events. In addition,
extra features have been developed and allow to mimic the response of an ideal detector
and to output the reconstructed events
into a simplified {\sc Lhco} or {\sc Lhe} format. Moreover, {\sc MadAnalysis}~5
is now also capable to generate DJR distributions allowing for the validation
of the jet merging procedure traditionally employed in state-of-the-art phenomenological
(and also in some experimental) analyses. 

\section*{Acknowledgments}
The authors are grateful to the ACAT organizers for the very nice conference.
They also thank A.~Alloul for his numerous checks of the code and for his help during the development
and the validation of the later releases.
This work has been supported by the French ANR 12 JS05 002 01 BATS@LHC and by the Theory-LHC France initiative
of the CNRS/IN2P3.

\section*{References}
\bibliographystyle{iopart-num}
\bibliography{ma5}
\end{document}